\documentclass[12pt]{article}
\usepackage{epsfig}

\begin{document}

\title{SySal: System of Salerno\footnote{Presented in The First International Workshop of Nuclear
Emulsion Techniques (12-14 June 1998, Nagoya, Japan),
http://flab.phys.nagoya-u.ac.jp/workshop.}}
\author{S. Amendola, E. Barbuto, C. Bozza, C. D'Apolito, \and A. Di Bartolomeo, M.
Funaro, G. Grella, G. Iovane, \and P. Pelosi and G. Romano. \\
\\
University of Salerno and INFN, Salerno, Italy}
\maketitle

\begin{abstract}
SySal (SYstem of SALerno) is the automatic scanning system developed in the
emulsion laboratory of Salerno to take part in the emulsion scanning phase
of the CHORUS experiment at CERN. In the following chapters we will present
features, results and further developments of this automatic multi-tracking
system.
\end{abstract}

\section{Introduction}

\noindent Since the early stages of high-energy physics, the detection of
sub-atomic particles and the study of their trajectories were often based on
visual techniques. The ''nuclear emulsion technique'' is a typical example:
charged particles crossing emulsion pellicles form a latent image as a trail
of sensitised silver bromide microcrystals. After development a three
dimensional ''track'' can be studied under high magnification by means of an
optical microscope.

Initially, human observers operated manually the microscope by moving the
stage, adjusting the focal plane of the objective, and examining a magnified
image through eyepieces. Later, the stage was motorised and the image made
available also on a TV screen; often the stage control and the image
analysis performed by the operator were assisted by a host computer, thus
configuring the so-called semi-automatic systems.

Ideally, the last step in this chain is to let the computer do the image
analysis usually performed by the operator to implement the instruments into
fully automatic systems.

SySal (SYstem of SALerno) \cite{rosa} is the automatic scanning
system developed in the emulsion laboratory of Salerno to take
part in the emulsion scanning phase of the CHORUS experiment at
CERN. In the following chapters we will present features, results
and further developments of this automatic multi-tracking system.

\section{SySal: A Multi-Tracking Scanning System}

\subsection{Hardware}

The hardware configuration of the system currently at work in Salerno
includes:

\begin{itemize}
\item  Nikon microscope with $40\times 40$ $cm^{2}$ stage, with $37$ $cm$
stroke on both horizontal directions, motor driven axes, $1$ $\mu m$
horizontal position accuracy, 0.5 $\mu $m vertical position accuracy;

\item  High-resolution ($1024\times 1024$ pixels) custom made CCD camera,
capable of $30$ frames per second at full resolution, and $60$ frames per
second in $512\times 1024$ mode (not used), with a minimum gate time of $0.9$
$ms$;

\item  Custom made stage controller with $3$ PID filters for the $3$ axes;

\item  Custom made lamp controller;

\item  Matrox Pulsar frame grabber card, with $4$ Mb on-board;

\item  PC Pentium MMX $233$ MHz with $64$ Mb RAM on-board, $4$ Gb HDD;

\item  Auxiliary monitor to see the image coming from the camera to the
frame grabber.
\end{itemize}

\bigskip

\subsection{The SYSAL Multi-Tracking Method}

The inspiring idea of SySal is very simple at first glance: basically, the
computer is asked to reproduce the human process of track recognition
presented in the introduction. The aim is recognizing tracks studying the
alignment of their grains through the emulsion layer.

Bearing this in mind one can use a computer to look at the emulsion at
different depths (usually $50$ levels for $350$ $\mu m$ thick emulsions) and
get tomographic images of all the tracks. These images must then be
assembled together for the real pattern recognition to take place. Once all
the tracks are reconstructed all the measurements are then easily obtained.

The figure 1 fixes the main steps of the analysis process.

\begin{figure}[h]
\begin{center}
\epsfig{file=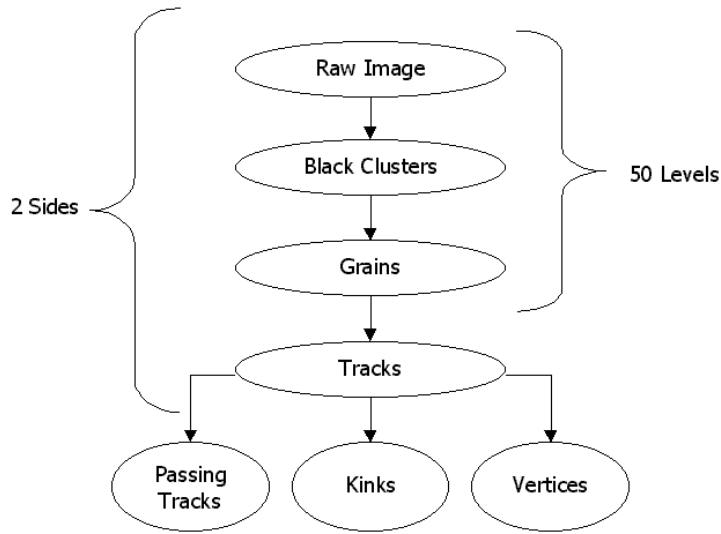,width=10cm,clip=}
\end{center}
\caption{Analysis Process}
\label{fig1}
\end{figure}

It is clear that, at the end of this process, the multi-tracking system
allows to get the whole physical information available (tracks, kinks and
vertices).

The scanning process must be smart enough to deal with ambiguous situations
and damaged information (this could come from zones with scratches or dirt
on the emulsion surfaces, and from inhomogeneous physical characteristics of
the plates); the main aim of an automatic system is, however, to do fast its
task, and this has not been forgotten. Most code has thus been written in
C++. When required, Assembler has also been used.

\subsubsection{Image Handling}

The starting point is, of course, the image coming from the camera. The
analog signal is digitized by the Pulsar, and converted to a gray scale of
256 levels (0 = black, 255 = white). The Pulsar stores the data while they
come into a 1 Mb internal linear buffer; however, this cannot be accessed
while it's receiving data. The double-buffering technique helps for the need
to have maximum speed: one 1 Mb buffer holds an image, making it available
to the CPU for analysis, while another buffer grabs the new image.

The CPU is in charge of recognizing black spots (called ''clusters'') in the
image: some of these are track grains in the emulsion; most clusters are
spurious grains, carrying no information, but physically existing in the
emulsion (they are called ''fog'' grains), due to the developing process;
some clusters come from noise in the electronic signal. Often the image of
the emulsion is not very clear due to shadows caused by grains that are not
in the focus plane or to scratches and dirt. Solving this problem is
possible using a simple FIR (Finite Impulse Response) filter. The result is
shown in figure 2.

\begin{figure}[h]
\begin{center}
\epsfig{file=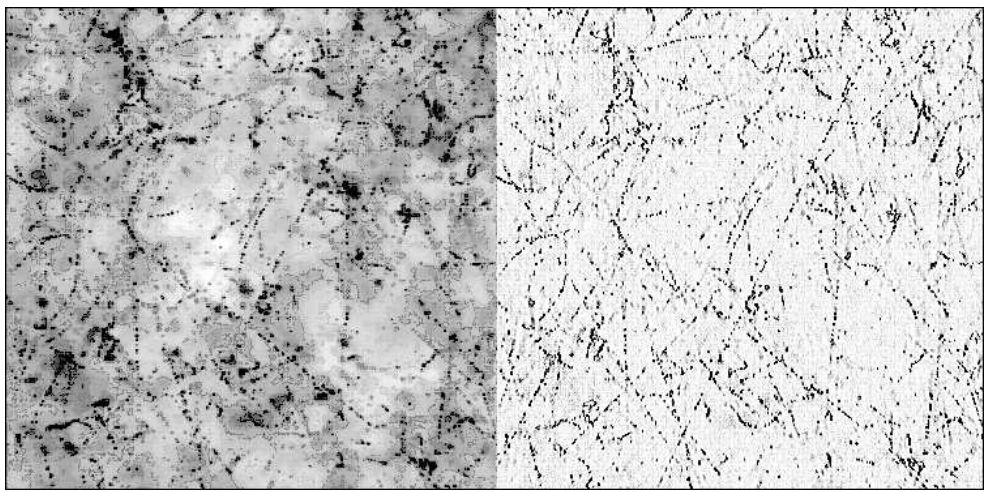,width=13cm,clip=}
\end{center}
\caption{On the left the image coming from the camera, on the right the FIR
filter is applied: wide black spots coming from electrons are resolved.}
\label{fig2}
\end{figure}

Basically, a well chosen threshold is used to look at the image, so to
divide the pixels in two classes: the ones with gray level above the
threshold (they are ''white'' pixels), and the ones with gray level below
the threshold (they are ''black'' pixels). The image stored in the buffer,
however, is not actually modified. The classification phase scans the image
row by row, from left to right. Each sequence of black pixels found is
called a ''segment'' and stored in memory. The segment finding process is
the most time-consuming, because it has to deal with a huge amount of data.
This led to write it in Assembler. After a row has been scanned, the new
segments are compared with the segments in the previous row; adjacent
segments are merged into a ''cluster'', an entity that has essentially an
area and a center. If two or more clusters come into contact, they are
merged.

\begin{figure}[h]
\begin{center}
\epsfig{file=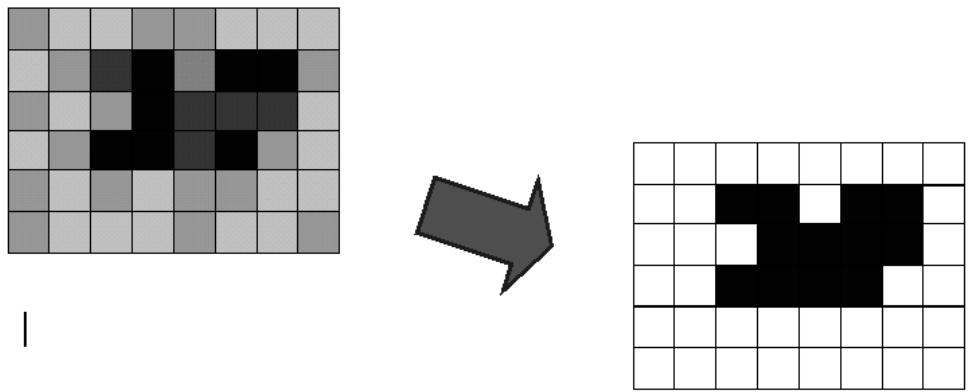,width=13cm,clip=}
\end{center}
\caption{The image is turned into black segments that are then assembled
into clusters.}
\label{fig3}
\end{figure}

\subsubsection{Track Recognition}

After all clusters have been formed, they are divided in three classes.
Small clusters are discarded (they mostly come from noise in the camera
signal); large clusters (sometimes called ''blobs'') are discarded too: they
are usually surface defects of the emulsion, or grains of heavy nuclear
fragments, usually with little energy, not so interesting for the
kinematical reconstruction of the interactions. Medium size grains survive,
and are used to recognize tracks left from high energy particles. These
should be straight; indeed, they are quite distorted, due to relaxation of
stresses in the emulsion during the developing process. Usually, straight
tracks are turned into parabolas, in such a way that the point lying at the
interface between the emulsion and its support remains in its original
position, and the slope of each track at its exit point in air is also left
unchanged (this may happen to be not exactly true, and some corrections can
be needed). A good multi-tracking algorithm must take into account this
phenomenon.

The tracking method is designed to reduce the computing time: to quote a
relevant number to understand the kind of job the CPU has to do, it can just
be said that, typically, for each of the $50$ layers, one can find $500$
grains out of $900$ clusters, and all of them are, in principle, grains that
may belong to a track.

\begin{figure}[h]
\begin{center}
\epsfig{file=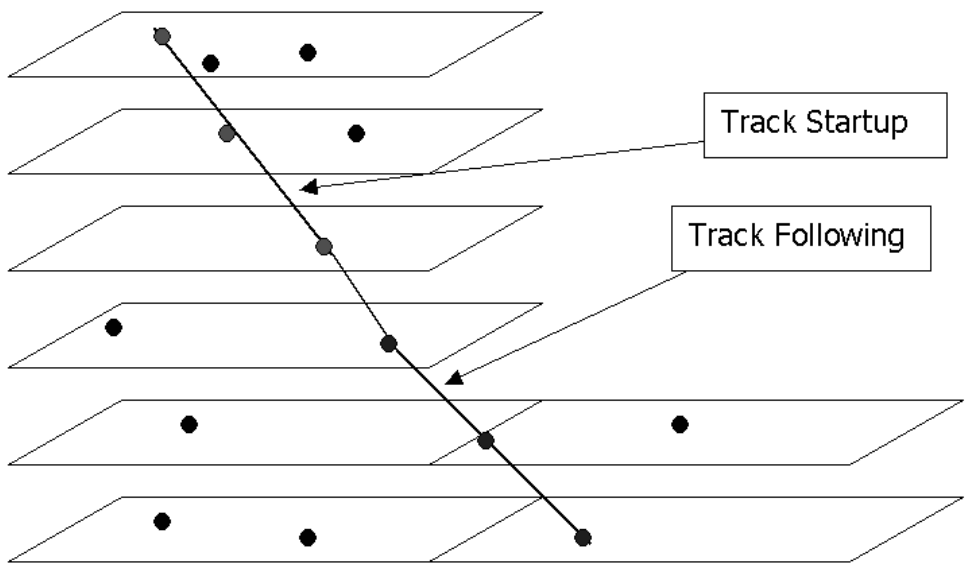,width=13cm,clip=}
\end{center}
\caption{The tracking procedure.}
\label{fig4}
\end{figure}

Each layer is divided in ''cells'', about $10\times 10$ $\mu m^{2}$ wide;
then, one looks for alignment of grains in $3$ adjacent layers, within cells
stacked one on top of the other. This phase is called ''track startup'':
when such an alignment is found, the computer looks for more aligned grains
in the next layers, up and down the startup point. During this ''track
following'' phase, the tracking algorithm is also allowed to change the cell
stack. After a new grain is appended, the current slope of the track is
recomputed; this makes it harder to miss next grains, because track
distortion is automatically accounted for. It may happen that, for some
reason, on one layer no grain is found belonging to the track being
followed. If this repeats for $5$ adjacent layers, the track following phase
ends.

While a track is being built, it may cross some already existing track. In
this case, if the tracks share three consecutive grains, they are joined
together in a single entity.

When a track stops both in the upward and downward directions, the number of
points collected is compared with a minimum threshold (usually $12$ points),
required to store the track in the final array.

\begin{figure}[h]
\begin{center}
\epsfig{file=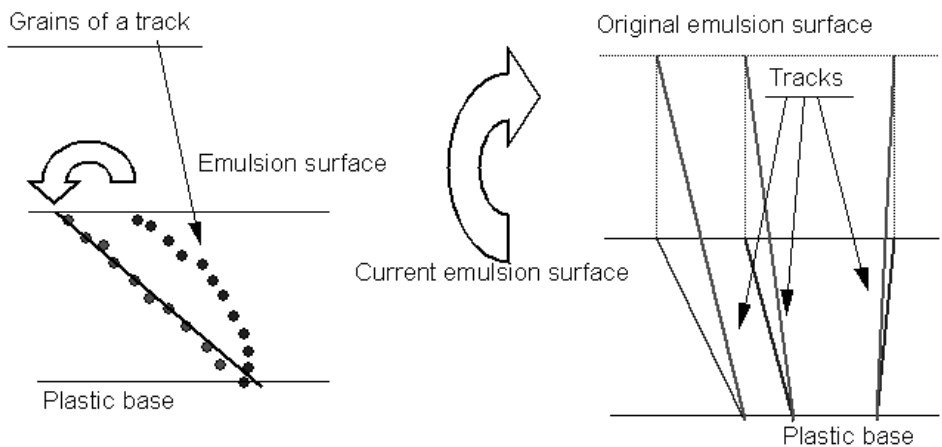,width=13cm,clip=}
\end{center}
\caption{Track corrections.}
\label{fig5}
\end{figure}

\subsubsection{Track Postprocessing}

After all the tracks in a field have been recognized (that's why we refer to
SySal as a multi-tracking system), the next step is to assign each of them
some global parameters, such as a slope and an intercept, obtained from a
linear fit. Of course, this cannot be done with distorted tracks. So, the
tracks that pass the whole thickness of the emulsion layer are used to
estimate the distortion vector, which is then used to correct the positions
of the grains of all the tracks in the current field. This correction relies
on the assumption that the slope at the exit point is the original one.
Sometimes this is not exactly true, and gives rise to some systematic
deviation of the computed slopes from the real values. However, this kind of
measurement error is not unrecoverable, and also an off-line correction is
possible.

Emulsions shrink during the developing process. So, when they are scanned,
the thickness is only $50\%$ of the original one. The reconstructed tracks
are thus finally ''expanded''.

The process described above is repeated for each of the two sides of the
emulsion. Then, for each track on the upper side, a pairing track on the
bottom side is searched, with the same angle and position (within proper
tolerances, typically $5$ $\mu m$ and $20$ $mrad$). When this process is
finished all the tracks in the scanning field are reconstructed without any
selection in slope so that the whole information present in the emulsion is
stored and available for the analysis.

\subsection{Brief  outlook  of  SySal  measurements in the \\ CHORUS experiment}

\begin{figure}[h]
\begin{center}
\epsfig{file=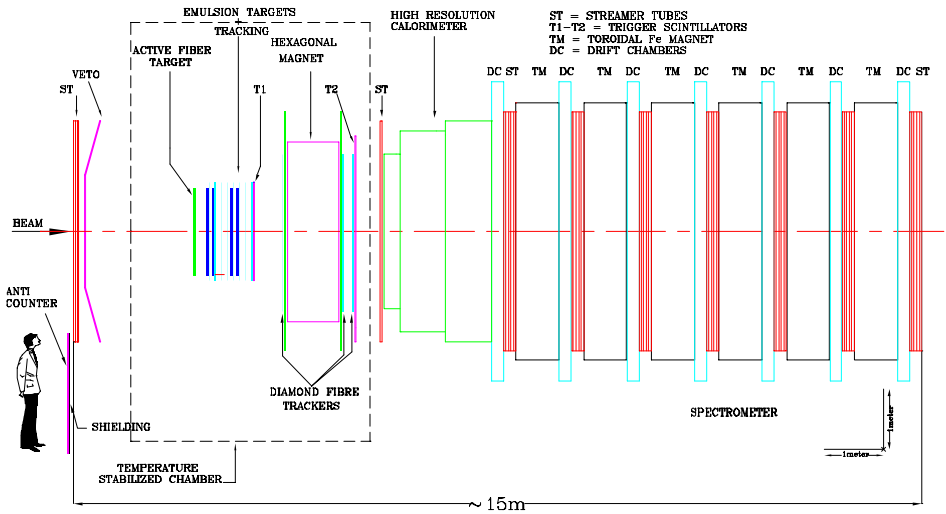,width=13cm,clip=}
\end{center}
\caption{Side view of the CHORUS detector.}
\label{fig6}
\end{figure}

Chorus (Cern Hybrid Oscillation Research apparatUS) \cite{chorus}
is a short baseline neutrino experiment. Its detector is installed
in the West Area of the CERN. It is composed of a bulk emulsion
target (four stacks of $36$ sheets of emulsion) interfaced through
some emulsion sheets (called special and changeable sheets) to the
electronic tracking part of the detector (scintillating fibers).
After these devices a magnetic spectrometer, a calorimeter and a
muonic detector follow (fig.6).

Scintillating fibers provide a prediction for every event that is therefore
followed back in the interface emulsion sheets. Then, using measurements in
these emulsion sheets as a new prediction, the tracks are followed back in
the target stack.

In the following some distributions of differences between predicted and
found coordinates (fig. 7, 8 and 9) and predicted and found slopes (fig. 10,
11 and 12) are presented for every kind of emulsion sheet used in the CHORUS
experiment.

\begin{figure}[hp]
\begin{center}
\epsfig{file=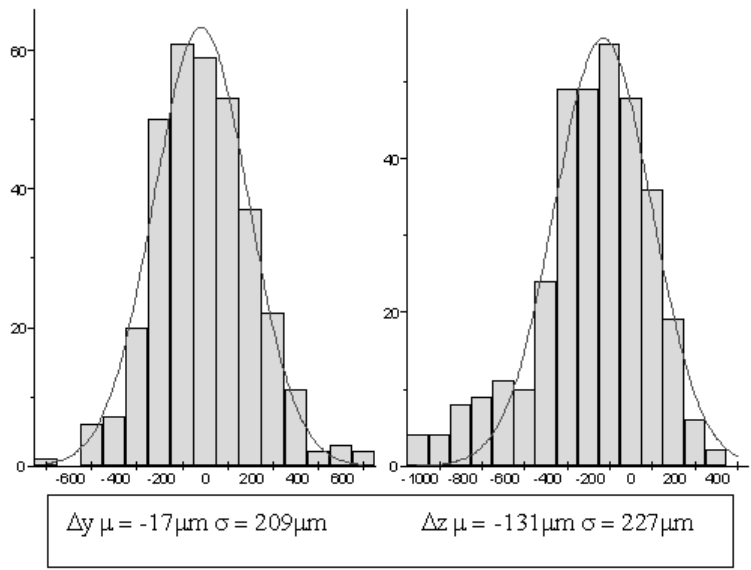,width=10cm,clip=}
\end{center}
\caption{Differences of predicted and found y (left) and z (right)
coordinates in changeable sheet.}
\label{fig7}
\end{figure}

\begin{figure}[hp]
\begin{center}
\epsfig{file=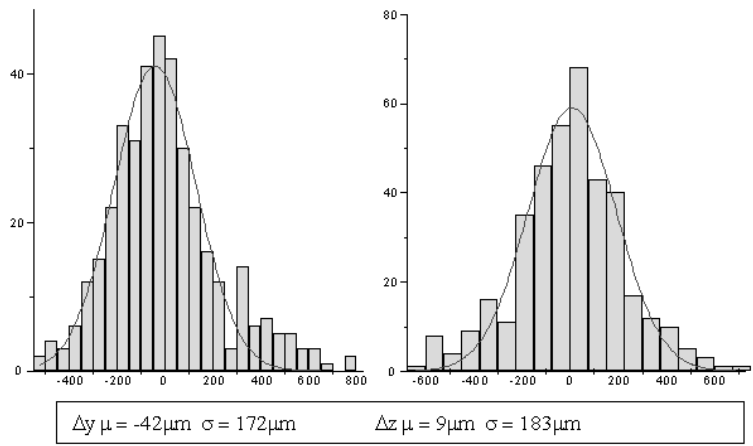,width=10cm,clip=}
\end{center}
\caption{Differences of predicted and found y (left) and z (right)
coordinates in special sheet.}
\label{fig8}
\end{figure}

\begin{figure}[hp]
\begin{center}
\epsfig{file=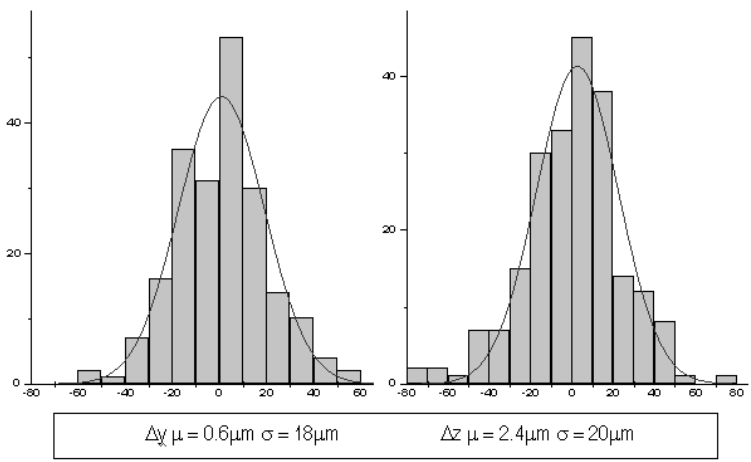,width=10cm,clip=}
\end{center}
\caption{Differences of predicted and found y (left) and z (right)
coordinates in bulk sheet.}
\label{fig9}
\end{figure}

\begin{figure}[hp]
\begin{center}
\epsfig{file=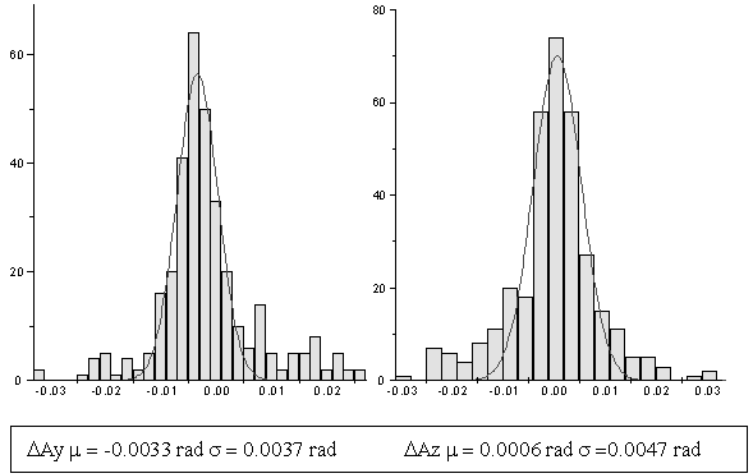,width=10cm,clip=}
\end{center}
\caption{Differences of predicted and found y (left) and z (right) slopes in
changeable sheet.}
\label{fig10}
\end{figure}

\begin{figure}[hp]
\begin{center}
\epsfig{file=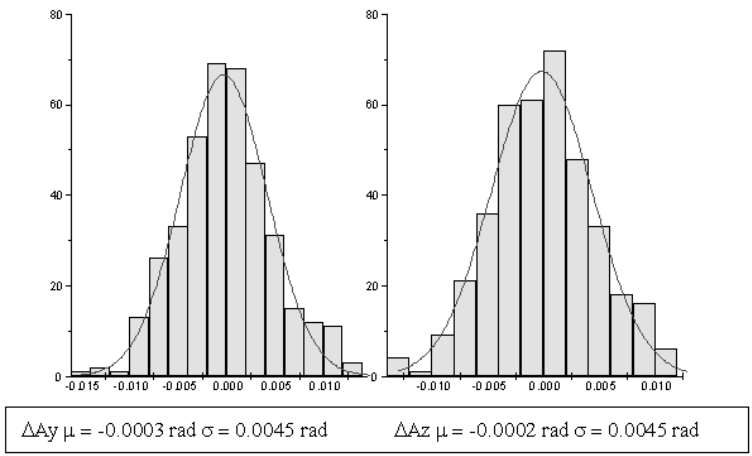,width=10cm,clip=}
\end{center}
\caption{Differences of predicted and found y (left) and z (right) slopes in
special sheet.}
\label{fig11}
\end{figure}

\begin{figure}[h!]
\begin{center}
\epsfig{file=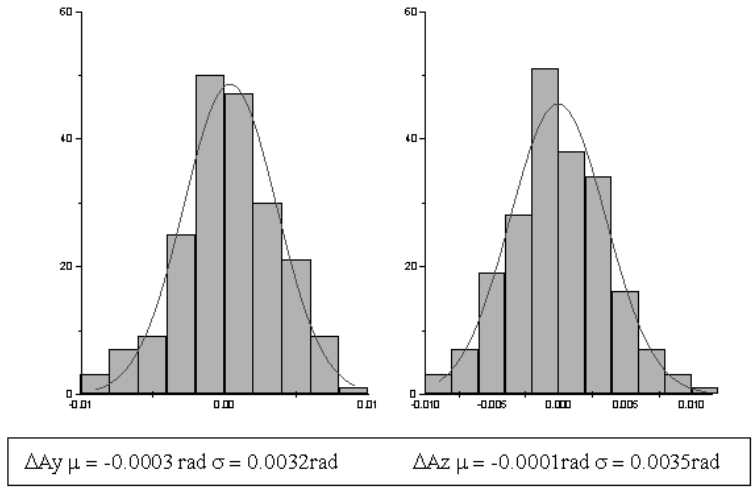,width=10cm,clip=}
\end{center}
\caption{Differences of predicted and found y (left) and z (right) slopes in
bulk sheet.}
\label{fig12}
\end{figure}

\section{Features and benefits of a Multi-Tracking System}

\subsection{The Multi-Tracking System}

As it turns out from the latter discussion, SySal is a multi-tracking
system. This means that during the scanning the system pays attention to all
the tracks in the field. Because of this, some important features are
stressed.

\begin{figure}[h!]
\begin{center}
\epsfig{file=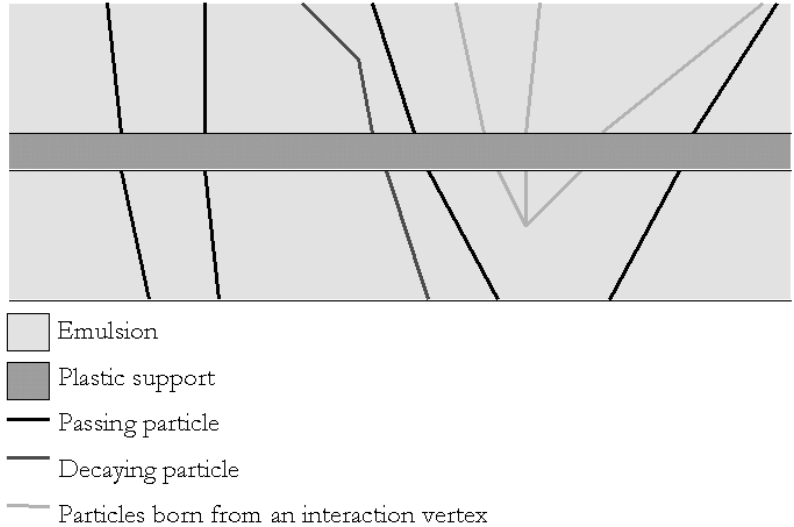,width=11cm,clip=}
\end{center}
\caption{Reconstructed tracks, with systematic errors on slope measurements
on the bottom side.}
\label{fig13}
\end{figure}

Reconstructing all the tracks rather than the single scanback track allows
the complete identification of an interaction vertex and of decay
topologies. Of course the physically interesting phenomenon is contained in
the interactions. If the particle momentum and energy are known, from the
geometrical parameters of a track before and after a kink one can get
information on the decaying particle. Using the kinematical data of the
daughter particles, and geometrical/topological knowledge of a vertex, the
complete kinematical reconstruction and the whole on-line study of an
interaction is therefore possible. The SySal approach is indeed the first
automatic scanning method that has been specifically designed for the task
of recognizing and reconstructing complete topologies.

Moreover, the local mapping of tracks near the scanback track can help for
the identification of the latter in another sheet of emulsion and can
provide information to improve the precision of angle measurement.

This is not all: mapping of wide zones of the emulsion is possible without
any further complication and can be successfully used for the
intercalibration of two sheets of emulsion. It follows that flux
measurements are very easy to perform. All these benefits are strictly
connected with the choice of a multi-tracking system.

\begin{figure}[h!]
\begin{center}
\epsfig{file=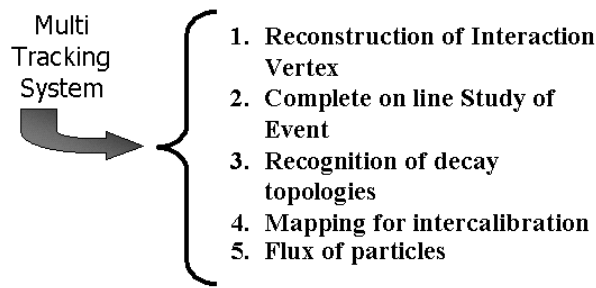,width=8cm,clip=}
\end{center}
\caption{Benefits of a Multi-Tracking System.}
\label{fig14}
\end{figure}

\subsubsection{Study of two interesting events}

As a sample of the features of a multi-tracking system, two interesting
events of the CHORUS experiment are presented. The good ability of SySal in
detecting kinks and in following back the tracks are stressed. Together with
this two requirements SySal satisfies the need for an on-line complete study
of the event.

The first interaction (Fig.15; Event 1339/380) was produced in the 1994 run
of the CHORUS experiment at CERN.

\begin{figure}[h]
\begin{center}
\epsfig{file=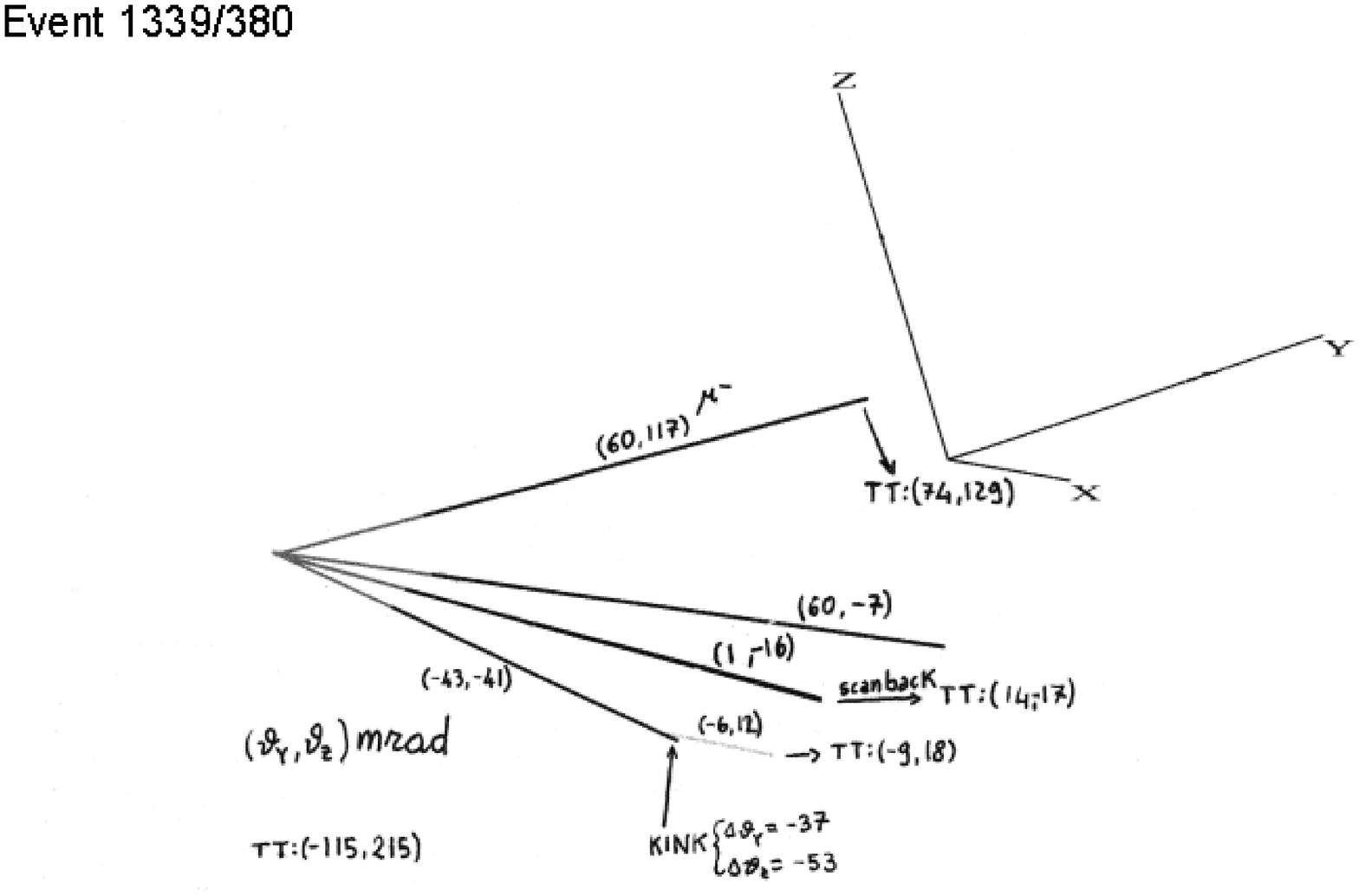,width=11cm,clip=}
\end{center}
\caption{Neutrino interaction shown in a 3D reconstruction by computer.}
\label{fig15}
\end{figure}

In this case, the interaction point was in the plastic base, $80$ $\mu m$
upstream of the nearest ends of the tracks in the emulsion; so the tracks
were extrapolated, and their extrapolations were found to cross.

In the following table we show the comparison between the prediction and the
real event, as found in the emulsion; angles are measured with respect to
the detector axis (X) at the vertex; the direction of a particle is defined
by the angles (YX,ZX).

\bigskip
\[
\begin{tabular}{|c|c|c|c|c|c|c|}
\hline
$Trk\#$ & $Pred.YX$ & $Pred.ZX$ & $FoundYX$ & $FoundZX$ & $Part.$ & $p\
GeV/c $ \\ \hline
$1$ & \multicolumn{1}{|r|}{$14\ mrad$} & \multicolumn{1}{|r|}{$-17\ mrad$} &
\multicolumn{1}{|r|}{$1\ mrad$} & \multicolumn{1}{|r|}{$-17\ mrad$} &
\multicolumn{1}{|r|}{$h^{-}$} & \multicolumn{1}{|r|}{$3.8$} \\ \hline
$2$ & \multicolumn{1}{|r|}{$74\ \,mrad$} & \multicolumn{1}{|r|}{$129\ mrad$}
& \multicolumn{1}{|r|}{$60\ mrad$} & \multicolumn{1}{|r|}{$117\ mrad$} &
\multicolumn{1}{|r|}{$\mu ^{-}$} & \multicolumn{1}{|r|}{$12.0$} \\ \hline
$3$ & \multicolumn{1}{|r|}{$-9\ mrad$} & \multicolumn{1}{|r|}{$-18\ mrad$} &
\multicolumn{1}{|r|}{$-43\ mrad$} & \multicolumn{1}{|r|}{$-41\ mrad$} &
\multicolumn{1}{|r|}{} & \multicolumn{1}{|r|}{} \\ \hline
$3\prime $ & \multicolumn{2}{|c|}{$After$ $380$ $\mu m$} &
\multicolumn{1}{|r|}{$-6\ mrad$} & \multicolumn{1}{|r|}{$12\ mrad$} &
\multicolumn{1}{|r|}{$h^{+}$} & \multicolumn{1}{|r|}{$5.4$} \\ \hline
$4$ & \multicolumn{1}{|r|}{} & \multicolumn{1}{|r|}{} & \multicolumn{1}{|r|}{%
$60\ mrad$} & \multicolumn{1}{|r|}{$-8\ mrad$} & \multicolumn{1}{|r|}{$?$} &
\multicolumn{1}{|r|}{$?$} \\ \hline
$5$ & \multicolumn{1}{|r|}{$-115\ mrad$} & \multicolumn{1}{|r|}{$215\ mrad$}
& \multicolumn{1}{|r|}{} & \multicolumn{1}{|r|}{} & \multicolumn{1}{|r|}{$?$}
& \multicolumn{1}{|r|}{$?$} \\ \hline
\end{tabular}
\]

\bigskip

Track 4 was not predicted, but was found at the vertex, while another track
was predicted and not found.

Track 3 was found in two pieces, with different slopes: this
sudden variation, after $380$ $\mu m$, of a high-momentum particle
cannot be due to scattering; so, one can conclude that there was a
decay. One can compute the transverse momentum of the daughter
particle with respect to the mother one,
by knowledge of the particle direction before the decay. The angles were ($%
-43$ $mrad$, $-41$ $mrad$). The variation of the particle direction is ($34$
$mrad$, $59$ $mrad$); the angle between the initial direction and the final
one is $68$ $mrad$.

The transverse momentum is

$p_{\perp }=p\sin \Delta \vartheta =5.4~GeV/c\times 0.068=367~MeV/c$

Information on the transverse momentum $p_{\perp }$ is useful to understand
what kind of decay we are analyzing. A $K^{+}$ decay into a $\pi ^{+}$ has a
maximum transverse momentum of $205~MeV/c$ so, taking into account our $%
p_{\perp }=367~MeV/c$, we must give up the hypothesis of such a decay. But a
$D^{+}$ decay into $\pi ^{+}$ has a maximum transverse momentum of $%
925~MeV/c $ and is therefore allowed.

This second hypothesis is in good agreement with the lifetime of the
particle that can be calculated from the length of its track in emulsion.

From the Lorentz transformation relating the momentum in the center of mass
with the momentum in the laboratory we yield:
\[
\left\langle \Delta \alpha \right\rangle \approx \frac{1}{\gamma }%
\Longrightarrow \gamma \approx 15
\]

Now the life time ($t^{\ast }$) of the decaying particle can be derived
using the relation:

\[
L=\beta \gamma ct^{\ast }
\]

Considering once again $\beta \approx 1$, we have:

\[
t^{\ast }\approx 10^{-13}s
\]

that is in good agreement with the charmed particle decay hypothesis.

Concluding we can give the following interpretation of the event:

\[
\begin{array}{c}
\nu _{\mu }N\rightarrow \mu ^{-}D^{+}X \\
D^{+}\rightarrow \pi ^{+}Y
\end{array}
\]

This study was very quick using the properties of the multi-tracking system
that gave us all the topological information related to the event during its
scanning. Then same basic relations of the high energy physics allowed us to
store and classify the event as a charm decay.

Let's analyze the second interesting interaction (Fig.16; Event 1495/371).

\begin{figure}[h]
\begin{center}
\epsfig{file=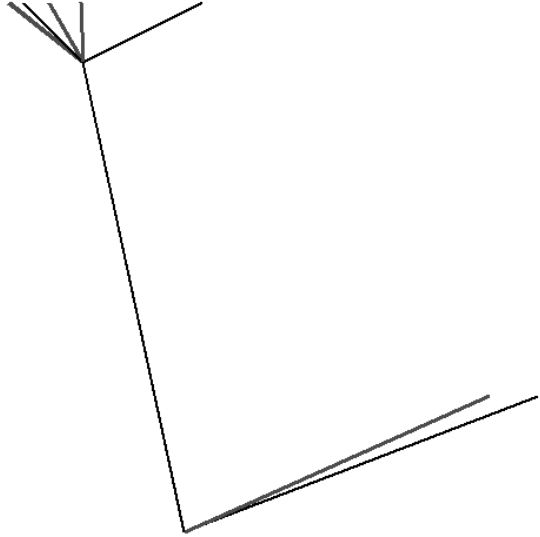,width=8cm,clip=}
\end{center}
\caption{The Interaction is shown in the xz plane.}
\label{fig16}
\end{figure}

This is a $0$ $\mu $ event with the primary interaction vertex in the $%
35^{th}$ bulk sheet and a secondary interaction in $33^{rd}$ bulk sheet. The
latter vertex had five tracks, three of them matching with the prediction
with a primary charged track. The scanning was updated and the old scanback
track ($Sy=38~mrad$,$Sz=13~mrad$) found at the secondary vertex was changed
into the primary charged track ($24~mrad$, $-36~mrad$) and followed back in
the previous sheet. In the $35^{th}$ sheet this track led to the primary
interaction vertex and so the topological reconstruction was complete.

Unfortunately the kinematics of the event was not completely reconstructed
by the CHORUS detector. At the secondary vertex two of the five
reconstructed tracks were not detected by the spectrometer and therefore
their charge and momentum are unknown. This makes impossible any analysis on
the primary charged track.

For the sake of completeness we must say that the secondary vertex could
signal a $5-prong$ decay of a $\tau $ that is however a very rare event. The
same can be said for a charmed particle, but identifying it as the
interaction of a secondary particle is more reasonable. The reconstruction
of this event was not problematic since the updating of the scanback track
is an automatic feature of the system.

\subsubsection{Intercalibration of two emulsion sheets}

A second sample of the ability of the multi-tracking system turns out in the
intercalibration procedure. In the CHORUS experiment, each one of the four
stacks is composed of 36 bulk sheets and so, in order to reach the
interaction vertex, the problem of getting all these sheets in the same
frame of reference has to be faced. One of the needs to get a successful
intercalibration between two bulk sheets is being able to match two maps of
tracks belonging to different sheets. For example tracks belonging to X7
beam (a muon beam passing trough the detector, useful to calibrate it) can
be followed. The procedure becomes harder taking into account that as a
first step the special sheet must be matched with the first bulk of the
stack. Due to the different time of exposure (1 year for the special sheet,
2 for the bulk sheet) about 50\% of the tracks in the bulk sheet are not
present in the special sheet. Matching tracks must be followed for 36 bulk
sheets, widening the scanning area because of the natural spread of the X7
beam.

The Multi-tracking system turns out to be very useful to build map of tracks
and to compare them, getting the transformation necessary to switch
coordinates of one map to the coordinates of the other.

Fig.17 shows two maps of tracks belonging to special and bulk (target)
sheets, in fig.18 these two maps are matched.

\begin{figure}[hp]
\begin{center}
\epsfig{file=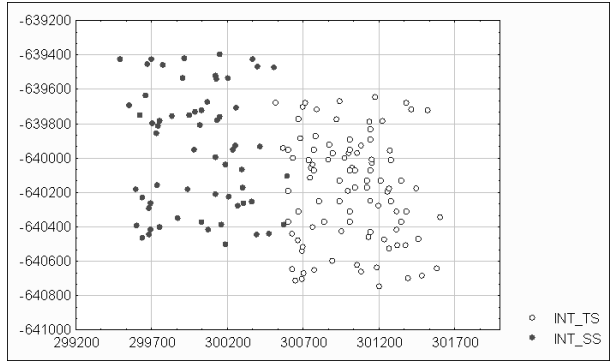,width=8cm,clip=}
\end{center}
\caption{On the left a special sheet map, on the right a bulk sheet map (in
the latter the number of tracks is almost double).}
\label{fig17}
\end{figure}

\begin{figure}[ph]
\begin{center}
\epsfig{file=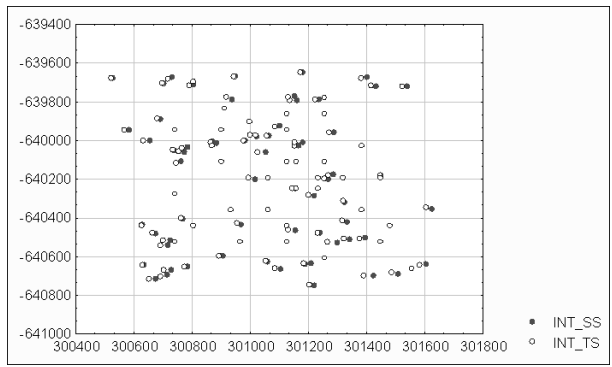,width=8cm,clip=}
\end{center}
\caption{Matching of the two previous maps of tracks.}
\label{fig18}
\end{figure}

\section{SySal approach to Vertex Detection: An ''Intelligent Scanning''}

\subsection{On-line vertex recognition and study in CHORUS experiment}

Vertex recognition and study using SySal is based on two important features.
The former is, of course, multi-tracking strategy and the latter is the
choice of following more than one track for every event.

During the scanning three vertex alarm are set:

\begin{itemize}
\item  Intersection of a scanback track in its scanning field with one or
more tracks matching with the event prediction;

\item  Disappearance of the scanback track;

\item  Closest approach of two scanback tracks belonging to the same event
(pre-location of the event).
\end{itemize}

Now we are going to discuss these three opportunities.

\subsubsection{Intersection of a scanback track in its scanning field with
one or more tracks matching with the event prediction}

Tracks matching with the scanback prediction are recognized in both side of
the emulsion. Intersections between these matching tracks and all the other
tracks in the scanning field are calculated; then a vertex alarm is given if
at least one of the other intersecting tracks matches with a track belonging
to the event prediction.

\begin{figure}[hp]
\begin{center}
\epsfig{file=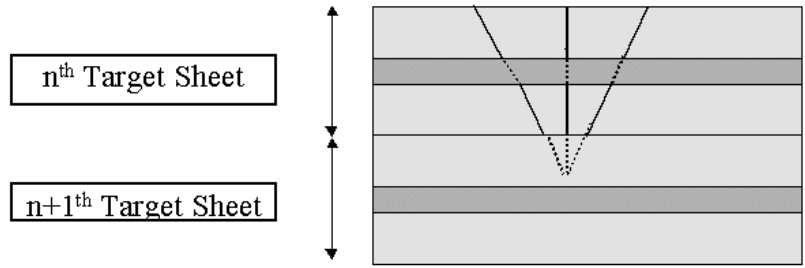,width=12cm,clip=}
\end{center}
\caption{Detection of the event is possible even from the sheet before the
vertex plate.}
\label{fig19}
\end{figure}

\begin{figure}[hp]
\begin{center}
\epsfig{file=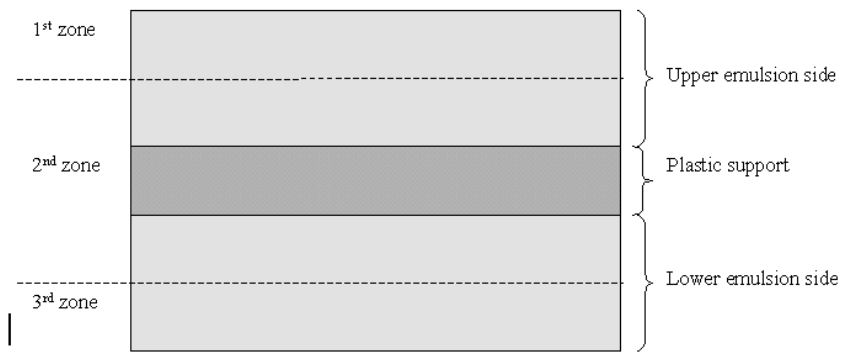,width=12cm,clip=}
\end{center}
\caption{Reconstruction of the event in the three zones of the vertex plate.}
\label{fig20}
\end{figure}

This alarm is able to recognize not only vertices in the current sheet of
emulsion but even in the following one (fig.19)

Vertex reconstruction and detection is somehow related to the local depth of
a vertex in the emulsion sheet. Therefore we can roughly divide the plate
thickness into three zones (fig.20).

Vertices located in the second zone of the emulsion sheet shown in fig.20
are usually well reconstructed in both sheets of emulsion (the vertex plate
and the previous one). Reconstruction of vertices in the first zone is
better in the previous bulk because in the vertex plate there are few grains
per track while vertices in the third zone are usually well reconstructed in
the vertex plate. For these vertices the other tracks belonging to the event
are often outside the scanning field of the previous bulk so that an
intersection is seldom possible.

Following (fig.21) there is an example of vertex found in bulk 9 but already
detected in bulk 8.

\begin{figure}[ph]
\begin{center}
\epsfig{file=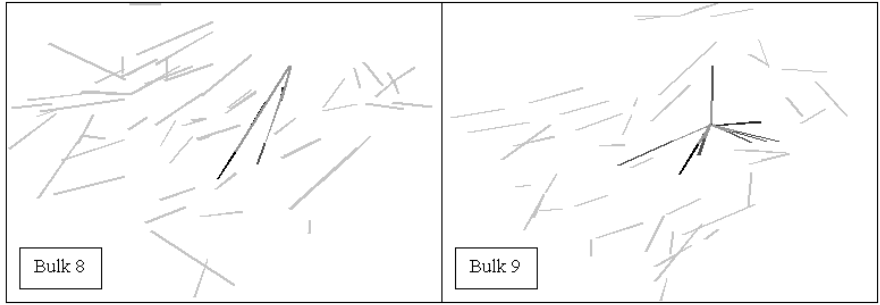,width=12cm,clip=}
\end{center}
\caption{An event in bulk 9 already detected in bulk 8.}
\label{fig21}
\end{figure}

In bulk 8 the scanback track is intersecting with a track matching with the
prediction. In the next bulk these two tracks are still intersecting and the
whole vertex is completely reconstructed.

\subsubsection{Disappearance of the scanback track}

Intersecting the scanback track with other tracks in the scanning field
won't be useful to detect quasi-elastic events. These events consist of only
one track predicted by the tracking system and so every intersection is
useless.

Therefore disappearing of the scanback track can be the winning strategy to
detect a kink of this track or the interaction vertex itself. Sometimes it
can happen that ''missing'' a linked track in a bulk sheet goes together
with the presence of a one side track matching with the prediction and
intersecting a wide angle track: this can be a further hint of the presence
of a vertex.

In general ''missing'' a track for two following bulk sheets can be used as
a vertex hint.

\subsubsection{Intersection of two scanback tracks: pre-location of the event%
}

This is an important matter because it can provide the chance of conceiving
a new scanning strategy. About this opportunity some ideas will be presented
in the next section.

Of course, following more than one track per event can be useful to locate
the event before reaching it, by calculating the intersection of two tracks
reconstructed in the same bulk sheet. To avoid any misleading information
coming from kinks it could be wise to follow even three tracks per event.

Let's now introduce the basics of such a study.

\textbf{Introducing the ''closest approach''}

Let's consider two points belonging to two different tracks in the space:

\[
\begin{array}{c}
P_{1}=(x_{1};\quad a_{1}^{y}x_{1}+b_{1}^{y};\quad a_{1}^{z}x_{1}+b_{1}^{z})
\\
P_{2}=(x_{2};\quad a_{2}^{y}x_{2}+b_{2}^{y};\quad a_{2}^{z}x_{2}+b_{2}^{z})
\end{array}
.
\]

The distance between them is:

\[
d=\sqrt{%
(x_{1}-x_{2})^{2}+(a_{1}^{y}x_{1}+b_{1}^{y}-a_{2}^{y}x_{2}-b_{2}^{y})^{2}+(a_{1}^{z}x_{1}+b_{1}^{z}-a_{2}^{z}x_{2}-b_{2}^{z})^{2}%
}.
\]

If we now consider only distances of points with the same x coordinate this
distance becomes:
\[
d=\sqrt{%
((a_{1}^{y}-a_{2}^{y})x+b_{1}^{y}-b_{2}^{y})^{2}+((a_{1}^{z}-a_{2}^{z})x+b_{1}^{z}-b_{2}^{z})^{2}%
}.
\]

Choosing the condition $\frac{\partial d}{\partial x}=0,$ we yield the depth
of the minimum distance (extrapolation depth):
\[
x_{\min }=-\frac{\left[ (a_{1}^{y}-a_{2}^{y})\left(
b_{1}^{y}-b_{2}^{y}\right) +(a_{1}^{z}-a_{2}^{z})\left(
b_{1}^{z}-b_{2}^{z}\right) \right] }{\left[
(a_{1}^{y}-a_{2}^{y})^{2}+(a_{1}^{z}-a_{2}^{z})^{2}\right] }
\]

\bigskip and so the minimum distance itself:
\[
d_{\min }=\sqrt{((a_{1}^{y}-a_{2}^{y})x_{\min
}+b_{1}^{y}-b_{2}^{y})^{2}+((a_{1}^{z}-a_{2}^{z})x_{\min
}+b_{1}^{z}-b_{2}^{z})^{2}},
\]

that is called ''closest approach''.

\textbf{A study of }$\mathbf{d}_{\mathbf{min}}$

This study was performed on about 50 closest approaches calculated during
the scanning of the stack 3 module 3 top part of the year 1994. These
closest approaches were derived from pairs of not kinking tracks that led to
interaction vertices. The aim is to check the precision of the measurements
of the vertex coordinates and the validity of the closest approach in
connection with the extrapolation depth.

As expected $d_{min}$ is strictly connected with the extrapolation
depth that is negative because the extrapolation is done backward
(fig.22).

\begin{figure}[h]
\begin{center}
\epsfig{file=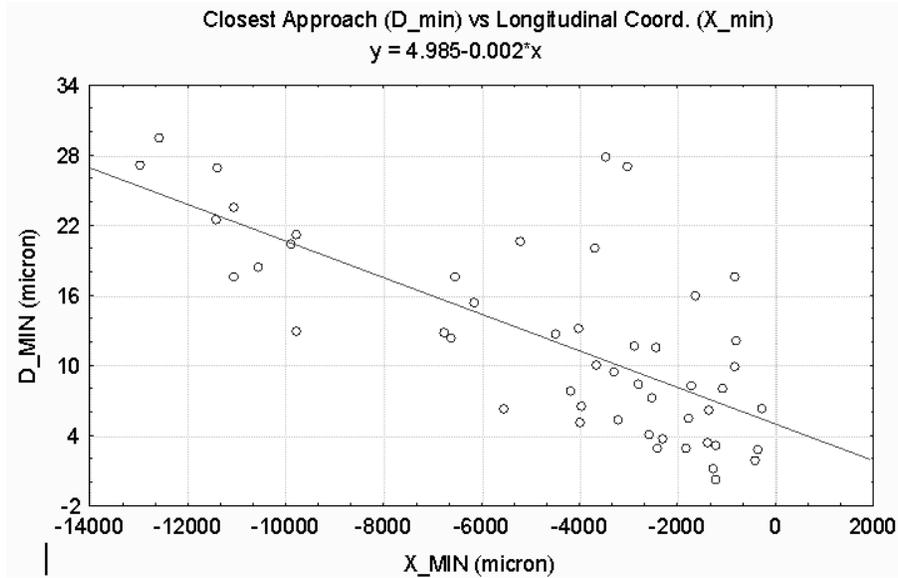,width=12cm,clip=}
\end{center}
\caption{Closest approach vs extrapolation depth.}
\label{fig22}
\end{figure}

Fig.23 shows the differences between the predicted (using
$d_{min}$) and the found transverse coordinates of the vertex. The
mean point of the segment representing the closest approach is
chosen as the predicted vertex. According to this, scanning of
only one field of view is necessary to locate the vertex using a
$50\times $magnification.

\begin{figure}[h!]
\begin{center}
\epsfig{file=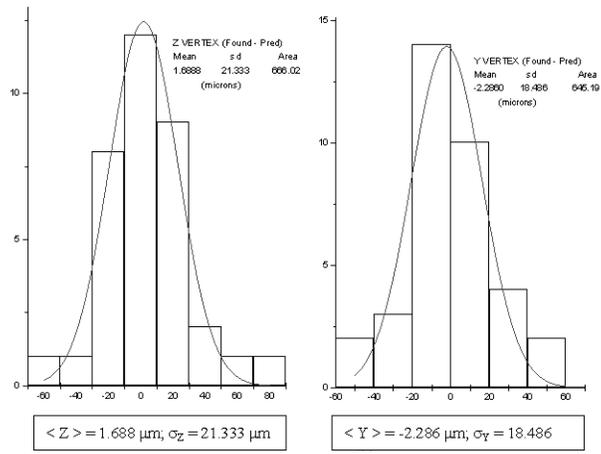,width=8cm,clip=}
\end{center}
\caption{Accuracy of the transverse coordinates of the vertex.}
\label{fig23}
\end{figure}

\begin{figure}[h!]
\begin{center}
\epsfig{file=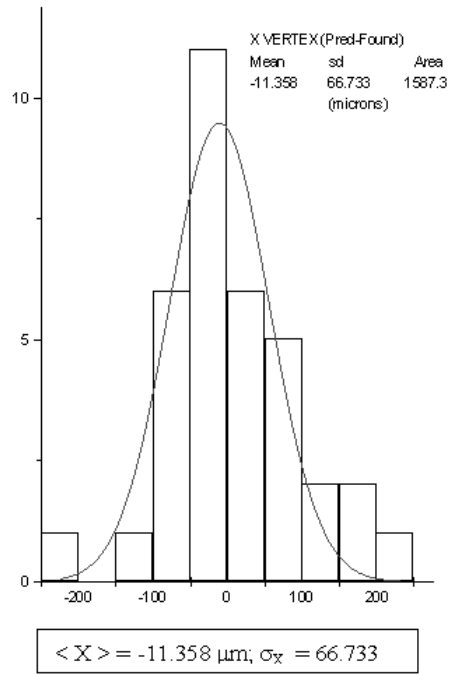,width=6cm,clip=}
\end{center}
\caption{Accuracy of the longitudinal coordinate of the vertex.}
\label{fig24}
\end{figure}

From the distribution of the differences of the longitudinal
coordinate (fig.24) it is easy to deduce that the interaction bulk
pellicle is well predicted.

We performed this study using tracks that led to vertices. Therefore this
study gives no suggestions about fake closest approaches, generated by pairs
of tracks in which there is a fake candidate. As it turns out from fig.22 $%
d_{min}$ increases with the extrapolation depth. So at long extrapolation
depths it is easy to mistake a good closest approach for a fake one and vice
versa.

\subsection{A further step: an ''intelligent'' scanning}

Time consumption is an important matter when dealing with huge amounts of
tracks to scan. That's why an ''intelligent'' scanning is required to skip:

\begin{itemize}
\item  all the fields that cannot be useful for the reconstruction of the
scanback track;

\item  all the sheets that would not give any important information about
interesting kink and vertex reconstruction.
\end{itemize}

\subsubsection{Skipping scanning fields}

The new version of SySal is considering the opportunity of skipping several
fields during the scanning of a track. This is performed using on-line
''intelligent'' choices.

\begin{figure}[h!]
\begin{center}
\epsfig{file=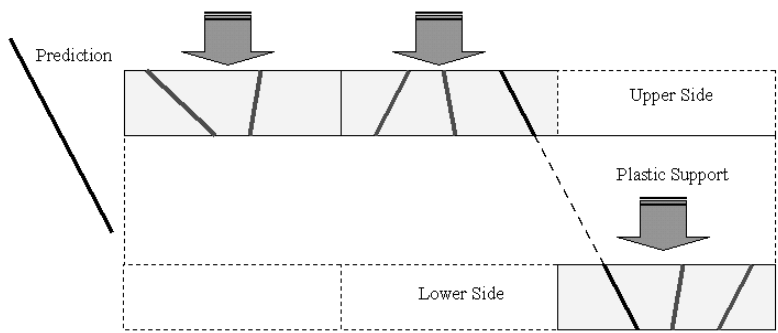,width=11cm,clip=}
\end{center}
\caption{When something interesting is found in the upper side a scanning is
performed in the lower side.}
\label{fig25}
\end{figure}

Whether in the upper side there is no track in good agreement with the
prediction then the lower side of the emulsion can be skipped; otherwise,
centering the interesting track and following down along its direction, a
scanning in the lower side of the emulsion can be performed.

Furthermore, scanning can be stopped if a good candidate (say within $5$ $%
mrad$) is found and the remnant part of the scanning area can be skipped.

\subsubsection{Skipping emulsion sheets}

The study of intersections of two tracks belonging to the same
event together with the information of the event prediction is
leading to the elaboration of a new scanning strategy. The goal is
to provide as soon as possible, with the minimum scanning effort,
a pair of tracks to locate the interaction vertex. This scanning
strategy is possible for those events with at least three ''safe''
tracks (well reconstructed by the detector, with angles
$<400~mrad$) otherwise a ''classical'' scanback scanning must be
used.

\begin{figure}[h!]
\begin{center}
\epsfig{file=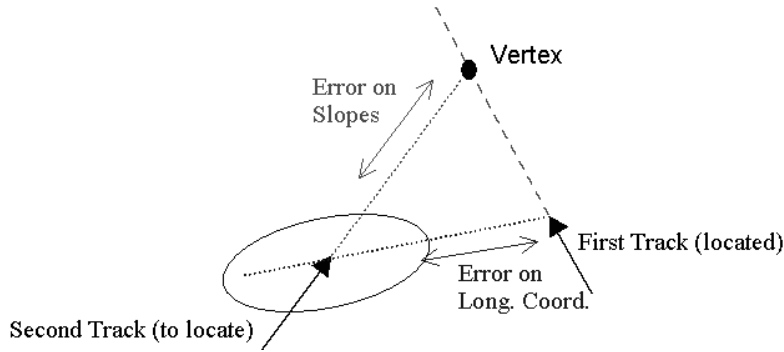,width=11cm,clip=}
\end{center}
\caption{Strategy to locate the second track.}
\label{fig26}
\end{figure}

Once the first track is located, using the predicted longitudinal coordinate
and the assumption that the vertex is along the direction of this track, it
is possible to locate the position of a second track. The second track must
be scanned in an area whose dimensions along the two directions in fig.26
are related with the uncertainty on the longitudinal vertex coordinate ($%
\sigma \simeq 1700~\mu m$) and on the predicted slopes of the tracks ($%
\sigma \simeq 3~mrad$).

When the second track is found and reconstructed it is possible to calculate
the closest approach otherwise the scanning of a third track can start.

Once two tracks are available their closest approach can be calculated. If
the extrapolation depth is too long it could be wise following the tracks
for the number of bulk sheets necessary to get a safe closest approach.
Otherwise one can jump directly to the bulk sheet necessary to begin the
analysis of the event (say a few sheets before the vertex plate).


\begin{thebibliography}{9}

\bibitem{rosa}
G.Rosa et al, Automatic Scan and Analysis of Digitized TV Images
by a Computer Driven Optical Microscope, NIM A 394 (1997) 357-367.

\bibitem{chorus}
 The CHORUS collaboration, The CHORUS experiment to search $\nu _{\mu
}\rightarrow \nu _{\tau }$ oscillation, CERN-PPE/97-33 March 14,
1997.

\end{thebibliography}
\end{document}